# BLEACHING OF OPTICAL ACTIVITY INDUCED BY UV LASER EXPOSURE IN NATURAL SILICA

M.Cannas* and F.Messina

*INFM and Dipartimento di Scienze Fisiche ed Astronomiche, Università di Palermo,*

*via Archirafi 36, I-90123*

*Palermo, Italy*

**Abstract**

We report experimental data on two types of natural silica, differing for their OH content, irradiated with UV photons (4.66 eV) from a pulsed Nd:YAG laser. Irradiation induces a reduction of the absorption band at 5.12eV and of the associated emissions at 3.14eV and 4.28eV, ascribed to twofold coordinated Ge (=Ge$^{\bullet\bullet}$) centers pre-existing in our samples. The bleaching is mainly due to the post-irradiation conversion of =Ge$^{\bullet\bullet}$ into the paramagnetic H(II) center via trapping of a H atom. Comparison with literature data points out the peculiarities of silica with a low Ge concentration as regards UV induced transformations.

* Corresponding author: M. Cannas

INFM and Dip.to di Scienze Fisiche ed Astronomiche, via Archirafi 36, I-90123 Palermo.

phone: +39 0916234220, fax: +390 916162461, e-mail: cannas@fisica.unipa.it

## 1. Introduction

In the framework of research pertaining to silica-based materials, UV-induced photosensitivity in Ge-containing $SiO_2$ is one of the issues that currently attracts more scientific interest. This is also motivated by the circumstance that photosensitivity can be exploited to build technologically relevant devices, such as fiber Bragg gratings [1] or optical memories [2].

In spite of the vast amount of work devoted in the last years to clarify the physical basis of $GeO_2:SiO_2$ photosensitive properties, a satisfactory understanding has yet to be reached. It has been supposed that a contribution to the change of optical features induced by UV radiation is due to point defects conversion phenomena, in which pre-existing centers transform in new defects after irradiation [3, 4]. In particular, a reduction of the ~5 eV absorption band upon UV exposure has been repeatedly observed and linked with photosensitivity. The decrease of this optical band is usually accompanied by the growth of a few characteristic electron paramagnetic resonance (EPR) Ge-related signals, depending also on sample chemical parameters, e.g. Ge content or hydrogen loading [3-6].

Many models have been introduced to explain the conversion processes transforming defects responsible for the ~5 eV absorption to EPR active centers. The puzzling picture of conversion models, emerging till now, is further complicated by the fact that at least two defects are currently debated to contribute to ~5eV absorption: the neutral oxygen vacancy, ≡Ge-Ge≡ or ≡Ge-Si≡, whose absorption band is peaked at 5.06 eV, and the twofold-coordinated Ge, =Ge•• (where each − stands for a bond with an O atom and • stands for an electron) responsible of absorption peaking at 5.16 eV and emissions at 3.1 eV and 4.2 eV [5, 7]. However, both the link between the two emissions and that between absorption at 5.16 eV and emissions have been recently questioned [8, 9].

Here we report an experimental study, performed using complementary optical and EPR spectroscopic methodologies, on UV-laser effects in natural silica specimens containing a very small (~1 ppm) Ge concentration. Our purpose is to investigate conversion processes on Ge-related centers in a different physical context than the one usually employed for technological issues and considered in several papers [3-9] (a few percent Ge concentration).

**2. Materials and Methods**

Our experiments were carried out at room temperature on 5x5x1 mm$^3$ samples of EQ906 and HERASIL1 natural silica. The former is a type I dry SiO$_2$ supplied by Quartz & Silice, while the latter is a type II wet produced by Heraeus Quartzglas. The two materials differ for OH content, it being 20 ppm by weight in EQ906 and 150 ppm in HERASIL1, as inferred from IR measurements on the OH stretching at 3600 cm$^{-1}$. Furthermore, from neutron activation measures we found the total concentration of Ge atoms: [Ge] = (1.4±0.3) ·10$^{16}$ cm$^{-3}$ in EQ906 and [Ge] = (1.6±0.3)·10$^{16}$ cm$^{-3}$ in HERASIL1.

UV-irradiations were performed with a pulsed Nd:YAG laser working in fourth harmonic mode (4.66 eV), emitting ~5 ns pulses of 15±1 mJ energy, uniformly distributed on a spot of 40±5 mm$^2$, with 1 Hz repetition frequency.

Optical absorption (OA) spectra in UV range were acquired with a JASCO V-570 double beam spectrophotometer, using a D$_2$ lamp as the light source and with 2 nm bandwidth.

Photoluminescence (PL) measures were carried out with a JASCO FP-770 spectrofluorimeter featuring a 150W Xe-lamp; all PL spectra reported in this work were obtained with a 5 nm excitation bandwidth and a 3 nm emission bandwidth and were corrected for spectral sensitivity of detecting system.

X-band EPR spectra were measured with a Bruker EMX spectrometer using a 100 kHz modulation. We adjusted the microwave power from P = 3.2 mW to P = 6.4 mW and the modulation amplitude from 0.2 mT to 0.4 mT to avoid saturation and distortion effects in revealing the induced paramagnetic centers. Absolute defect concentrations were obtained from comparison of double integrated EPR spectra with those from a reference silica sample whose defect concentration, E′ centers, was measured by spin-echo decay method [10]; absolute accuracy is 20%.

**3. Results**

To study the effects induced by Nd:YAG laser, we performed 6 irradiation experiments for each of the two examined materials, with a number of laser shots ranging from 20 to 3000 in wet silica samples and from 100 to 10000 for dry ones.

Many days after each experiment, when post-irradiation annealing processes were completed [11], we measured PL spectra under 5.0 eV excitation in our samples, comparing them with those observed before irradiation. Results are reported in Fig. 1 for wet HERASIL1 silica. All spectra show two emission bands centered respectively at 3.14±0.01 eV with full width at half maximum FWMH = 0.42±0.02 eV and 4.28±0.01 eV, FWMH= 0.46±0.03 eV. The main evidence emerging from these data is the progressive reduction of PL intensity on increasing the number of laser shots. We stress that UV irradiation induces a change of the ratio between the intensities of the emission bands at 3.14 eV and 4.28 eV: this ratio, measured to be 5.1±0.4 before irradiation, increases of ~30 % after 3000 laser pulses. In the inset, we report the effect produced by laser exposure on absorption spectrum where one observes a partial bleaching of a band peaked at 5.12±0.02 eV, FWHM= 0.45±0.03 eV ($B_{2\beta}$ band [12]).

These bleaching effects induced by UV rays on optical activity of HERASIL1 silica are better evidenced in Fig. 2 where we show the total luminescence intensity (sum of 3.14 eV and 4.28 eV band areas) and the $B_{2\beta}$ band amplitude as a function of the number of laser shots. PL and OA signals decrease in agreement as irradiation dose grows from 20 to 3000 shots, their final value being ~(40-50)% of the initial one. An analogous bleaching effect was also observed in the dry samples EQ906 by measuring the PL intensities, which are shown in the inset of Fig. 2 with irradiation dose; reduction is ~40% over 10000 laser shots. We observe that in these samples the reduction of the $B_{2\beta}$ absorption band is partially concealed by the simultaneous growth of the band at 5.8 eV associated with the Si-E' ($\equiv$Si•) centers [13].

To clarify the conversion processes associated with the bleaching of optical bands, we used EPR spectroscopy to investigate the UV induced paramagnetic defects. In Fig. 3 we show a composite spectrum comparing EPR signals observed in the two investigated silica materials. The dry EQ906 sample was irradiated with 200 laser shots, the wet HERASIL1 with 330 shots; the EPR spectra were detected many days after the laser exposure. The growth of several components due to different centers is clearly visible. In EQ906, we observe a doublet with 11.8 mT splitting due to H(II) centers (=Ge$^\bullet$-H) [14, 15], whose concentration is found to be $C=(8\pm2)\times10^{14}$ cm$^{-3}$. In the central part of the spectrum, one observes the saturated and distorted line of Si-E′ centers and a weak signal visible on the right of Si-E′ resonance line. By the comparison with data from literature [16], we can ascribe the latter structure to the Ge(2) center, consisting in a hole trapped on a twofold coordinated Ge (=Ge$^\bullet$)$^+$ [6]. However, their partial superposition with the Si-E' center signal prevents us from quantitatively evaluating their concentration. At variance in the HERASIL1 sample only the H(II) centers signal is clearly visible, their concentration being $C=(2.2\pm0.4)\times10^{15}$ cm$^{-3}$.

**4. Discussion**

UV irradiation with a Nd:YAG laser on natural silica samples produces two effects:

a) Bleaching of the absorption band at 5.12 eV ($B_{2\beta}$ band) and of the two emissions at 3.14 eV and 4.28 eV, associated with twofold coordinated Ge center [16]. On increasing the number of laser shots, the reduction of the two optical activities, OA and PL, is quite similar.

b) Growth of two Ge-related paramagnetic centers, H(II) and Ge(2), the latter observed only in dry samples.

The first point gives evidence of a conversion of twofold coordinated Ge induced by UV photons. The substantial agreement between OA and PL signals indicates that the overall optical activity arises from the same center, as previously put forward by experimental and theoretical works [17, 18]; a strong point in favor of the single center model is the linear correlation between the optical bands found in a large number of as grown natural silica samples [19]. This is in contrast with recent works on X-irradiated Ge-doped silica [8] and UV irradiated Ge-doped optical fiber preforms [9], both showing a lack of correlation among these optical bands. We point out that the linear correlation between OA and PL bands involves a constant luminescence quantum yield, while that between the two emissions is based on a constant intersystem crossing rate linking the excited singlet and triplet states from which 4.28 eV and 3.14 eV emissions take place, respectively [17-19]. Hence, a departure from correlation can arise from a variation of these parameters induced by a selective conversion of centers whose features are inhomogeneously distributed [19-21]. This is consistent with the observed change of the ratio between the emission intensities at 3.14 eV and 4.28 eV evidenced in our results.

The second point allows us to address the transformation processes of the optically active defect upon laser UV irradiation. The growth of H(II) centers, observed in both silica types, is associated with the conversion of the twofold coordinated Ge by trapping of an hydrogen atom [14, 15, 22]:

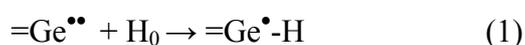

$$=\text{Ge}^{\bullet\bullet} + \text{H}_0 \rightarrow =\text{Ge}^{\bullet}\text{-H} \qquad (1)$$

Recent experimental work on the same samples demonstrated the occurrence of post-irradiation phenomena related to diffusion of mobile hydrogen [11]. In particular, the formation of H(II) centers mostly happens within a few hours after the end of irradiation and the hydrogen available for the reaction (1) is produced by radiolysis of O-H or Si-H bonds.

The observation of Ge(2) centers in dry samples gives evidence of a further conversion channel of =Ge$^{\bullet\bullet}$ in this silica type. In agreement with previous studies [6], this process takes place via the UV photo-ionization of the twofold coordinated Ge. The absence of Ge(2) defects in our wet samples could be due to their annealing by reaction with hydrogen, whose concentration in form of O-H bonds is higher, by about an order of magnitude, than in dry silica. Hence, the so produced Ge(2)-H groups could become H(II) centers by trapping an electron [23]. This process may be a further contribution to the post irradiation growth of H(II) centers observed in our samples [11]. To verify this hypothesis a more complete study on Ge(2) centers annealing is needed.

As a final remark, we comment on the contrast between our results and the response to UV rays of heavy Ge doped silica [4-7, 23]. In fact, we did not observe either Ge(1) $(GeO_4)^-$ [7] or Ge-E′ ($\equiv$Ge•) [5] centers; so we hypothesize that this is due to the lower concentration of fourfold or threefold coordinated Ge precursors in our natural silica specimens than in Ge-doped ones. As a proof, the concentration of twofold coordinated Ge in the as grown EQ906 and HERASIL1 was measured to be $[=Ge^{••}]\sim 7\times 10^{15}$ cm$^{-3}$ [22], which compared with the total Ge concentration, $[Ge]\sim 1.5\times 10^{16}$ cm$^{-3}$, demonstrates that the ratio $[=Ge^{••}]/[Ge]\sim 0.5$ is much higher than in the heavy Ge-doped samples, $[=Ge^{••}]/[Ge]\sim 10^{-4}$-$10^{-2}$ [7, 24].

## 4. Conclusions

We investigated conversion processes of Ge-related defects in natural silica samples with a very low Ge concentration (~1 ppm). The main observation is the reduction of the absorption and luminescence bands associated with the twofold coordinated Ge defect after UV irradiation with a Nd:YAG laser. This bleaching is accompanied by formation of two paramagnetic Ge-related centers: H(II) and Ge(2). The prevalent process is the conversion of $=Ge^{••}$ into H(II) due to diffusion limited reactions with mobile hydrogen occurring after the laser irradiation. These results, compared with literature data on Ge-doped silica, point out the relevance of Ge content in determining the conversion channels activated by UV rays.

**Acknowledgements**

Stimulating discussions with R. Boscaino, S. Agnello, F.M. Gelardi and M. Leone are gratefully acknowledged. The authors thank G. Lapis and G. Napoli for technical assistance. This work was partially supported by a National Project (PRIN2002) of the Ministero dell'Istruzione, dell'Università e della Ricerca, Rome (Italy).

**References**

[1] K.O.Hill, Y. Fujii, D.C. Johnson, B.S. Kawasaki, Appl.Phys.Lett. 32 (1978) 647.

[2] M. Watanabe, S. Juodkazis, H.-B. Sun, S. Matsuo, H. Misawa, Appl. Phys. Lett . 77 (2000) 13.

[3] G. Pacchioni, L. Skuja, D. Griscom (Eds.), Defects in $SiO_2$ and Related Dielectrics: Science and Technology, Kluwer Academic Publishers, Dordrecht, 2000.

[4] Neustrev, J. Phys.: Condensed Matter 6 (1994) 6901.

[5] J. Nishii, K. Fukumi, H. Yamanaka, K. Kawamura, H. Hosono, H. Kawazoe, Phys. Rev. B 52 (1995) 1661.

[6] M. Fujimaki, T. Watanabe, T. Katoh, T. Kasahara, N. Miyazaki, Y. Ohki, H. Nishikawa, Phys. Rev. B 57 (1998) 3920.

[7] H. Hosono, Y. Abe, D. Kinser, R. Weeks, K. Muta, H, Kawazoe, Phys. Rev. B 46 (1992) 11445.

[8] D.P. Poulios, J.P. Spoonhower, N.P. Bigelow, J. Lumin. 101 (2003) 23.

[9] B. Poumellec, M. Douay, J.C. Krupa, J. Garapon, P. Niay, J. Non Cryst. Solids 317 (2003) 319.

[10] S. Agnello, R. Boscaino, M. Cannas, F.M. Gelardi, Phys. Rev. B, 64 (2001) 1946.

[11] M. Cannas, S. Agnello, R. Boscaino, S. Costa, F.M. Gelardi, F. Messina, J. Non Cryst. Solids, 322 (2003) 90.

[12] R. Tohmon, H. Mizuno, Y. Ohki, K. Sasagane, K. Nagasawa, Y. Hama, Phys. Rev. B 39 (1989) 1337.

[13] R.A. Weeks, C.M. Nelson, J. Appl. Phys. 31 (1960) 1555.

[14] J. Vitko, J. Appl. Phys. 49 (1978) 3511.

[15] V.A. Radtsig, A.A. Bobyshev, Phys. Stat. Sol. (b) 133 (1986) 621.

[16] E.J. Friebele, D.L. Griscom, G.H. Sigel Jr., J. Appl. Phys. 45 (1974) 3424.

[17] L. Skuja, J. Non Cryst. Solids 149 (1992) 77.

[18] B.L. Zhang, K. Raghavachari, Phys. Rev. B, 55 (1997) R15993.

[19] M.Leone, S.Agnello, R. Boscaino, M. Cannas, F.M. Gelardi. Phys Rev. B 60 (1999) 11475.

[20] V.N. Bagratashvili, S.I. Tsypina, V.A. Radtsig, A.O. Rybaltovskii, P.V. Chernov, S.S. Alimpiev, Y.O. Simanovskii, J. Non Cryst. Solids 180 (1995) 221.

[21] L. Skuja, J. Non Cryst. Solids 239 (1998) 16.

[22] S. Agnello, R. Boscaino, M. Cannas, F.M. Gelardi, M. Leone, Phys. Rev. B 61 (2000) 1946.

[23] M. Fujimaki, T. Kasahara, S. Shimoto, N. Miyazaki, S. Tokuhiro, K. Seol, Y. Ohki, Phys. Rev. B 60 (1999) 4682.

[24] S. Grandi, P.C. Mustarelli, S. Agnello, M. Cannas, A. Cannizzo, J. Sol-Gel Sci. Technol. 26 (2003) 915.

**Figure captions:**

**Figure 1**) PL spectra detected in wet silica HERASIL1 under 5.0 eV excitation before and after exposure to different doses of Nd:YAG laser radiation. In the inset, OA spectra of the same samples are reported.

**Figure 2**) Emission intensity (measured as sum of the two band areas under 5.0 eV excitation) and absorption coefficient at 5.1 eV with number of laser shots for wet HERASIL1 $SiO_2$. In the inset, emission intensity detected in dry samples EQ906 is shown.

**Figure 3**) (A) Composite EPR spectrum showing the main signals observed in dry EQ906 silica irradiated with 200 laser shots; the inset shows a simulation of Ge(2) signal taken from Ref. [16]. (B) EPR spectrum of the wet HERASIL1 sample exposed to 330 laser shots.

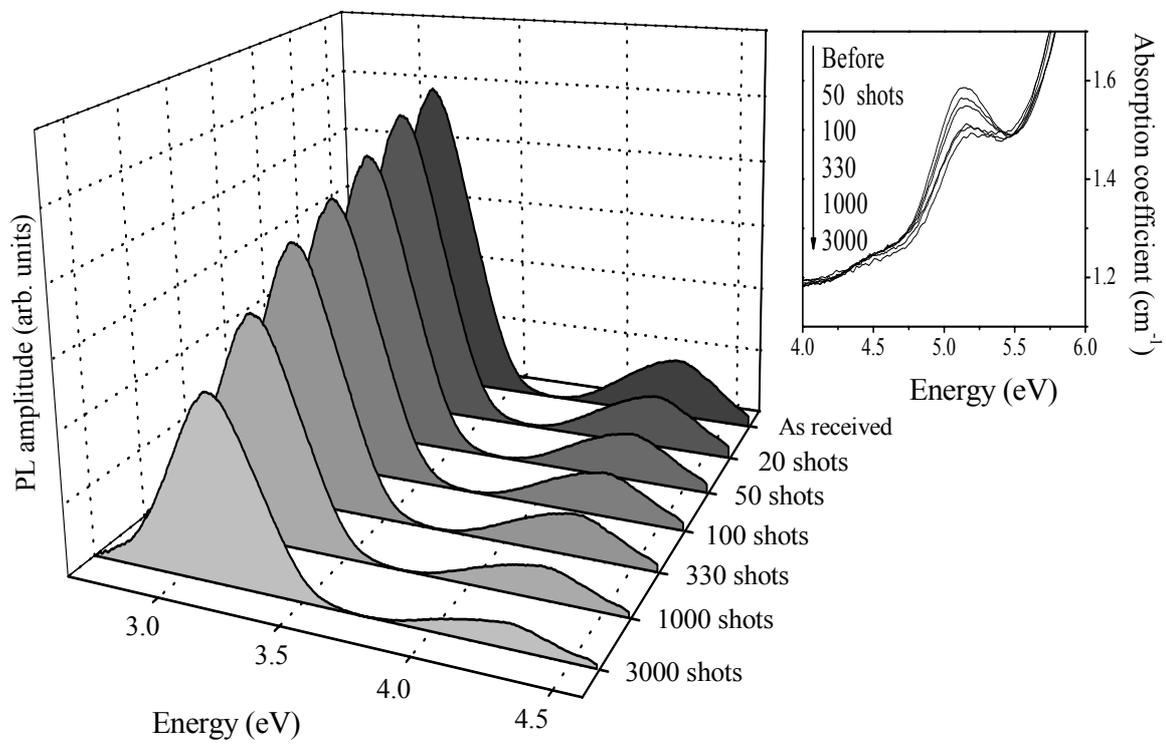

**Figure 1**

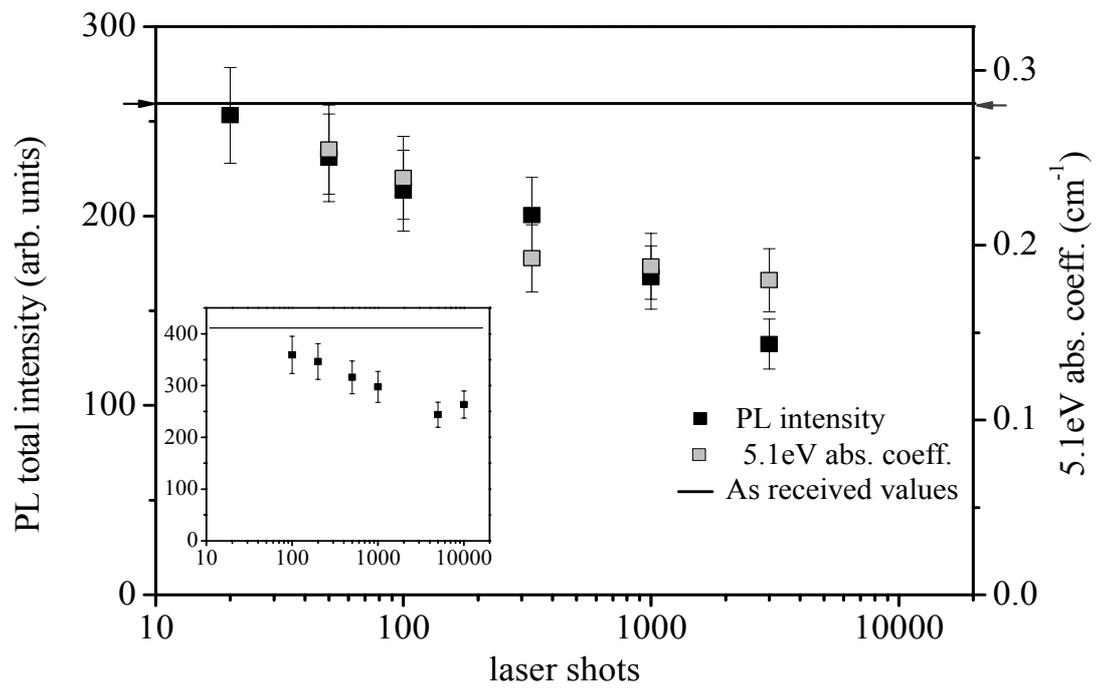

**Figure 2**

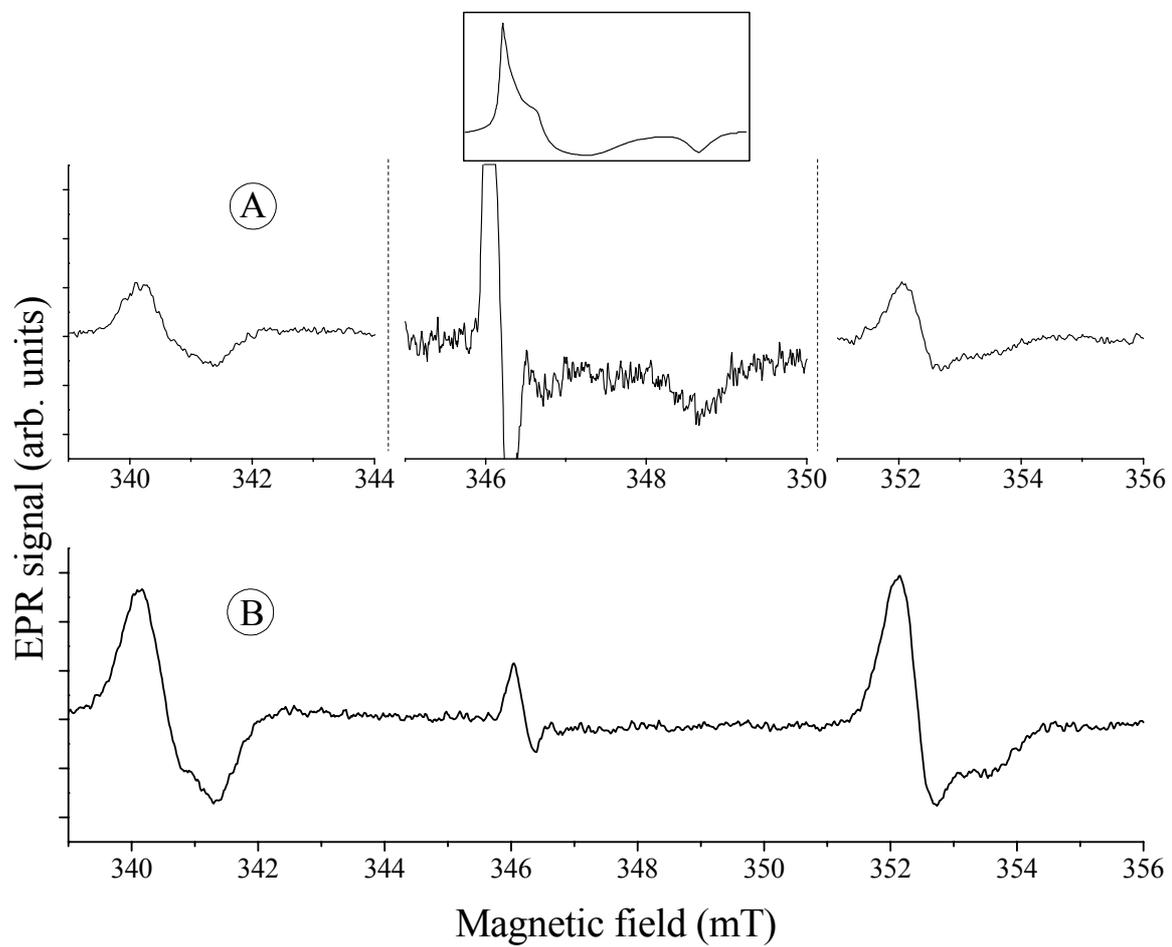

**Figure 3**